\documentclass[12pt]{article}
\usepackage{graphicx,subfigure,psfrag}
\usepackage{amsfonts,amsmath}

\usepackage{color}
\usepackage{tabularx}
\newcommand{\pc}[1]{\parbox{0pt}{#1}}
\usepackage[colorlinks=false]{hyperref} 

\usepackage{dsfont}
\usepackage{times}

\newcommand{\R}{{\mathbb R}}
\newcommand{\Z}{{\mathbb Z}}

\def\+{{+\!\!\!+}}

\newcommand{\C}{{\mathds C}}

\def\pmb#1{\setbox0=\hbox{#1}%
\kern.0em\copy0\kern-\wd0 
\kern-.04em\copy0\kern-\wd0 
\kern.08em\copy0\kern-\wd0 
\kern-.04em\raise.0433em\box0 }         
 
\newcommand{\nc}{\newcommand} 
\nc{\beq}{\begin{equation}} 
\nc{\eeq}[1]{\label{#1}\end{equation}} 
\nc{\ber}{\begin{eqnarray}} 
\nc{\eer}[1]{\label{#1}\end{eqnarray}} 
\nc{\pek}[1]{\cite{#1}} 
\nc{\enr}[1]{(\ref{#1})} 
\nc{\kal}[1]{{\cal{#1}}} 
\nc{\dott}{\;\cdot\;} 

\def\0 {\nonumber}

\begin{document} 

\setcounter{page}{0}
\newcommand{\inv}[1]{{#1}^{-1}} 
\renewcommand{\theequation}{\thesection.\arabic{equation}} 
\newcommand{\be}{\begin{equation}} 
\newcommand{\ee}{\end{equation}} 
\newcommand{\bea}{\begin{eqnarray}} 
\newcommand{\eea}{\end{eqnarray}} 
\newcommand{\re}[1]{(\ref{#1})} 
\newcommand{\qv}{\quad ,} 
\newcommand{\qp}{\quad .} 

\def\qp{Q_+}
\def\qm{Q_-}
\def\qbp{\bar Q_+}
\def\qbm{\bar Q_-}
\def\sgh{\Sigma_{g,h}}

\begin{titlepage} 
\begin{center} 

                         
\vskip .3in \noindent 


{\Large {\bf Refined Wall-Crossing, Free Fermions and Crystal Melting}}


\vskip .3in 

\end{center}

\centerline{Haitao Liu $^{1}$, and Jie Yang$^2$}
\medskip{\baselineskip14pt
\centerline{$^1$ Department of Mathematics and Statistics, University of New Brunswick, Fredericton, Canada}
\centerline{\it and}
\centerline{Theoretical Physics Group, The Blackett Laboratory, Imperial College London, London, UK}
\centerline{\it and}
\centerline{Department of Applied Mathematics, Hebei University of Technology, Tianjin, China}
\centerline{\it haitao.liu@unb.ca}
\vskip .1in
\centerline{$^2$ School of Mathematical Sciences, Capital Normal University, Beijing, China}
\centerline{\it and}
\centerline{SISSA, INFN, Sezione di Trieste, via Bonomea 265, Trieste, Italy}
\centerline{\it yang9602@gmail.com}
                   }

\vskip 1in

\begin{center} {\bf ABSTRACT }  
\end{center} 
\begin{quotation}\noindent  
In this paper, we use an M-theory model to conjecture the refined reminiscence of the 
OSV formula connecting the refined topological string partition function with the refined BPS states partition function for the toric Calabi-Yau threefolds without any compact four cycles. Further, we show how to use the vertex operators in 2d free fermions to reproduce the refined BPS states partition function for the $\mathbb{C}^3$ case and the wall-crossing formulas of the refined BPS states partition function for the resolved conifold and $\mathcal{O}(-2)\oplus\mathcal{O}\rightarrow \mathbb{P}^1$ cases. 

\end{quotation} 
\vfill 
\eject

\end{titlepage}

\tableofcontents

\section{Introduction}

Recent years people have made lots of progress on understanding the space of BPS states, $\mathcal{H}_{BPS}$, in type II string compactifications on Calabi-Yau threefolds. In general, such compactifications give rise to the effective $\mathcal{N}=2$ theories in four dimensions. $\mathcal{H}_{BPS}$ is a special subspace of the full Hilbert space which is the one-particle representation of the $d=4, \mathcal{N}=2$ supersymmetry algebra. In fact, the space $\mathcal{H}_{BPS}$ contains lots of information about the Calabi-Yau threefold $X$. It is a bridge connecting the black hole physics and topological strings \cite{Ooguri:2004zv}. 

In fact, the space $\mathcal{H}_{BPS}$ is graded by charge sectors
\begin{equation}
\mathcal{H}_{BPS}=\oplus_{\gamma\in\Gamma}\mathcal{H}_{BPS}(\gamma),
\end{equation}
where $\Gamma$ denotes the charge lattice. According to \cite{Harvey:1996gc, Minasian:1997mm, Denef2007} in the Calabi-Yau compactification of type IIA string theory, the lattice $\Gamma$ is given by the cohomology of the Calabi-Yau threefold $\Gamma=\text{H}^{\text{even}}(X; \mathbb{Z})$, and $\gamma$ is given by the generalized Mukai vector of the stable coherent sheaves corresponding to the D6/D4/D2/D0 branes
\begin{equation}
\begin{array}{ccccccccc}
\gamma=ch(\mathcal{E})\sqrt{\hat{A}(X)}&=&p^0&+&P&+&Q&+&q_0\\
&\in& \text{H}^0&\oplus&\text{H}^2&\oplus&\text{H}^4&\oplus &\text{H}^6\\
& & D6 & &D4& &D2& &D0
\end{array}
\end{equation}
Further, $\mathcal{H}_{BPS}(\gamma)$ has the following decomposition due to the existence of the universal hypermultiplets 
\begin{equation}
\mathcal{H}_{BPS}(\gamma)=(\mathbf{0},\mathbf{0};\mathbf{\frac12})\otimes \mathcal{H}'_{BPS}(\gamma).
\end{equation}
In this paper we will focus on the reduced space $\mathcal{H}'_{BPS}$. It is well known that the space $\mathcal{H}'_{BPS}$ depends on the asymptotic boundary conditions in the four-dimensional spacetime, where the boundary conditions in IIA compactification are the complexified K\"{a}hler moduli $u=iJ+B$ of the Calabi-Yau threefold $X$ \cite{Denef2007}. Roughly speaking, $\mathcal{H}'_{BPS}(\gamma, u)\cong \text{H}^*(\mathcal{M}(\gamma,u))$, where $\mathcal{M}(\gamma,u)$ is the moduli space of the stable coherent sheaf with the generalized Mukai vector $\gamma$ under certain $u$-dependent stability condition \cite{Diaconescu2007}. Further, on $\mathcal{H}_{BPS}$ there is an Spin(3) action, which gives rise to the following index of the $\mathcal{H}'_{BPS}$ \cite{Dabholkar2005, Denef2007}
\begin{equation}
\Omega(\gamma, u):={\rm Tr}_{{\cal H}_{BPS}'(\gamma,u)} (-1)^{2J_3'},\footnote{Here we have factorized the contribution of the universal hypermultiplets \cite{Denef2007}.}
\end{equation}
where $J_3'$ is the reduced angular momentum \cite{Denef2007}. From the mathematical point of view, the index $\Omega(\gamma, u)$ is related to the Euler characteristic of the space $\mathcal{H}'_{BPS}(\gamma,u)$. The interesting phenomenon is that the index $\Omega(\gamma, u)$ is piecewisely constant with respect to $u$. In other words, on the moduli space $u$ there are some chambers which split the moduli space into some small chambers. When we enter a new chamber crossing a wall, the index $\Omega(\gamma, u)$ will jump to another integer. The detailed wall-crossing formulas of $\Omega(\gamma, u)$ have been well understood for many cases \cite{Denef2007,Szendroi2007, Gaiotto2010a, Young2008, Aganagic2009,Aganagic2010} etc. Notice that the index $\Omega(\gamma, u)$ is only related to the Euler characteristic of the space $\mathcal{H}'_{BPS}(\gamma,u)$. Therefore we want to know how the cohomology jumps when we cross a wall. In \cite{Dimofte2010a}, the authors defined the following \emph{refined} index 
\begin{equation}
\Omega^{ref}(\gamma, u, y):={\rm Tr}_{{\cal H}'_{BPS(\gamma,u)}} (-y)^{2J_3'},
\end{equation}
where $y$ is a free parameter. Hence $\Omega(\gamma, u, y)$ is a polynomial of $y$. Thus we have
\begin{equation}
\Omega^{ref}(\gamma, u, y)\rightarrow\Omega(\gamma,u)\ \ \text{ as } \ y\rightarrow 1.
\end{equation}
In \cite{Dimofte2010a}, the authors claimed that the \emph{refined} index $\Omega^{ref}(\gamma, u, y)$ is related to the q-deformed Motivic BPS index of Kontsevich and Soibelman in \cite{Kontsevich2008} via the following map
\begin{equation}
\begin{array}{ccccccccc}
\text{Refined}& &\text{Motivic}&&\text{Quantum}\\
y&\leftrightarrow &\mathbb{L}^{\frac12}&\leftrightarrow&-q^{\frac12}
\end{array}
\label{eqintro1}
\end{equation}
They conjectured that in string theory on a Calabi-Yau threefold $X$, the q-deformed motivic index of Kontsevich and Soibelman in \cite{Kontsevich2008} corresponds to turning on a graviphoton background on $\mathbb{R}^4$. In particular, we have 
\begin{equation}
\Omega^{ref}(\gamma,u,y)=\Omega^{mot}(\gamma,u)
\end{equation}
with the appropriate identification of variables $y, \mathbb{L}, q$.

Further, like the unrefined index we can define the refined BPS states partition function \cite{Dimofte2010a} by
\begin{equation}
\mathcal{Z}^{ref}_{BPS}(q,Q,y,u):=\sum_{\substack{\beta\in\text{H}_2(X;\mathbb{Z})n\in\mathbb{Z}}}(-q)^nQ^\beta\Omega^{ref}(\gamma_{\beta,n},u,y).
\footnote{Here the meaning of $q$ is not the same as the $q$ in the eq. (\ref{eqintro1}). We use the same notation $q$ here just for making the expression of the partition function consistent with other papers.}
\end{equation}
In \cite{Dimofte2010a}, the author have shown how to use the dimer model to produce the wall-crossing formula of the refined BPS states partition function of the resolved conifold case. In this paper, we show how to use the vertex operators in 2d free fermions and the crystal corresponding to the Calabi-Yau threefold $X$ to reproduce the wall-crossing formula of the refined BPS states partition function. 

\noindent{\emph{Organization of the paper}}

In section \ref{sec:ref_OSV}, we conjecture the \emph{refined} reminiscence of the 
OSV formula by using the M theory model; in section \ref{sec:fermion}, we review some basic knowledge of the vertex operators in the two-dimensional free fermions; in section \ref{sec:C3}, we show how to use the vertex operators to produce the refined BPS states partition function on ${\mathbb C}^3$; in section \ref{sec:conifold}, we show how to use the vertex operators to reproduce the wall-crossing formulas of the refined BPS states partition function on the resolved conifold; in section \ref{sec:Ominus2}, we show how to use the vertex operators to reproduce the wall-crossing formulas of the refined BPS states partition function on the $\mathcal{O}(-2)\oplus\mathcal{O}\rightarrow\mathbb{P}^1$; in section \ref{sec:discussion}, we discuss some possible future working directions.

\section{Refined partition function of BPS states and refined reminiscence of the OSV formula}
\label{sec:ref_OSV}

In \cite{Aganagic2009}, the authors employ the Taub-Nut space with the unit charge to lift the single D6 brane in type IIA theory to M-theory.  Since the degeneracy of BPS states does not depend on the radius of the Taub-Nut circle, we can investigate the index of the BPS states in the large radius case, and then go back to the small radius case. Now let us focus on the large radius case in which the Taub-Nut becomes $\R^4$. In other words, our 11-dimensional spacetime of M-theory is 
\begin{equation}
X\times \R^4\times S^1,
\label{eq1}
\end{equation}
where $X$ is a Calabi-Yau threefold and $S^1$ is the time circle. Here we take the following assumptions which are the same as those in \cite{Aganagic2009}:
\begin{enumerate}
\item{Assume that the K\"{a}hler parameters of the Calabi-Yau are vanishing.}
\item{Assume that the only BPS states in 5D are particles.}
\end{enumerate}
The assumption 1 implies that the central charges of the BPS M2 branes are parallel; the assumption 2 implies that there are not any compact four-cycles in the Calabi-Yau threefold $X$ \cite{Aganagic2009}. Therefore from now on when we use the symbol $X$, it always refers to a Calabi-Yau threefold without any compact four-cycles.

Now let us consider the physics in the five-dimensional spacetime $\R^4\times S^1$. Since the isometry group of $\R^4$ is $SO(4)=SU(2)_L\times SU(2)_R$, we have two kinds of internal spins: left one $m_1$ and right one $m_2$. The $U(1)_L\times U(1)_R\subset SU(2)_L\times SU(2)_R$ action on $\R^4=\C^2$ is as follows
\begin{eqnarray}
U(1)_L&:&(z_1, z_2) \mapsto (e^{i\epsilon_1}z_1, z_2), \\
U(1)_R&:&(z_1, z_2) \mapsto (z_1, e^{i\epsilon_2}z_2),
\end{eqnarray}
where $z_1, z_2$ are the complex coordinates of $\C^2$ and $e^{i\epsilon_1}, e^{i\epsilon_2}$ are the generators of $U(1)_L, U(1)_R$ actions respectively. 
Further, each five-dimensional BPS particle can have excitations on $\R^4=\C^2$. These excitations can be expressed by holomorphic functions on $\C^2$ \cite{Hollowood2008, Aganagic2009}:
\begin{equation}
\Phi_{\beta, m_1, m_2}(z_1, z_2)=z_1^{m_1}z_2^{m_2}\sum_{n_1, n_2\geq 0}\alpha_{\beta, n_1+m_1, n_2+m_2}z_1^{n_1}z_2^{n_2},
\label{eq2}
\end{equation}
where $z_1, z_2$ are the coordinates on $\C^2$, $\beta\in H^2(X, \Z)$ is a curve class wrapped by the M2-brane, and $\alpha_{n_1+m_1, n_2+m_2}$ are bosonic or fermionic modes depending on whether the field $\Phi$ is bosonic or fermionic respectively. Denote by $N_{\beta}^{j_L, j_R}$ the degeneracy of the five-dimensional BPS states of M2 branes of charge $\beta$ and left-right spin $(j_L, j_R)$ under $SU(2)_L\times SU(2)_R$. Then for each mode $\alpha_{\beta, n_1+m_1, n_2+m_2}$ we have
\begin{eqnarray}
2j_L^3&=&n_1+n_2+m_1+m_2, \label{eq5} \\
2j_R^3&=&n_1-n_2+m_1-m_2. \label{eq6}
\end{eqnarray}

Following \cite{Hollowood2008, Iqbal2009, Aganagic2009}, we may define the following unrestricted refined partition function: 
\begin{eqnarray}
&&{\cal Z}_{Fock}=\text{Tr}_{Fock}(-1)^{2(j_L+j_R)}q_1^{j_L^3+j_R^3}q_2^{j_L^3-j_R^3}e^{-T}\nonumber\\
&=&\prod_{\beta\in H_2(X, {\mathbb Z})}\prod_{j_L,j_R}\prod_{k_L=-j_L}^{+j_L}\prod_{k_R=-j_R}^{+j_R}\prod_{m_1,m_2=1}^\infty(1-q_1^{k_L+k_R+m_1-\frac{1}{2}}q_2^{k_L-k_R+m_2-\frac{1}{2}}Q^\beta)^{(-1)^{2(j_L+j_R)+1}N_\beta^{j_L,j_R}},\nonumber\\
\label{eq7}
\end{eqnarray}
where $q_1=e^{i\epsilon_1}, q_2=e^{-i\epsilon_2}, Q=e^{-T}$ and $T$ is the area of $\mathbb{P}^1$.
According to \cite{Hollowood2008, Iqbal2009, Behrend2009}  the refined topological string partition function is defined by 
\begin{eqnarray}
&&{\cal Z}_{top}^{ref}=\mathcal{M}(q_1, q_2)\text{Tr}_{\cal
  H}(-1)^{2(j_L+j_R)}q_1^{j_L^3+j_R^3}q_2^{j_L^3-j_R^3}e^{-T}=\mathcal{M}(q_1, q_2)\cdot\nonumber\\
&&\cdot\prod_{\substack{\beta\in H_2(X, {\mathbb Z})\\
\beta>0}}\prod_{j_L,j_R}\prod_{k_L=-j_L}^{+j_L}\prod_{k_R=-j_R}^{+j_R}\prod_{m_1,m_2=1}^\infty(1-q_1^{k_L+k_R+m_1-\frac{1}{2}}q_2^{k_L-k_R+m_2-\frac{1}{2}}Q^\beta)^{(-1)^{2(j_L+j_R)+1}N_\beta^{j_L,j_R}},\footnotemark\nonumber\\
\label{eq8}
\end{eqnarray}
\footnotetext{When we prepare this paper, we notice Behrend et al.'s paper \cite{Behrend2009}. In their paper, they claim that $\mathcal{M}(q_1, q_2)^2=M_{\delta_1}(q_1,q_2)\cdots M_{\delta_{\chi(X)}}(q_1, q_2)$ is related to the Poincar\'{e} polynomial of the Hilbert scheme of points on the Calabi-Yau threefold. Here we postpone the discussion on $M_{\delta_1}(q_1,q_2)\cdots M_{\delta_{\chi(X)}}(q_1, q_2)$ to the section \ref{sec:discussion} in this paper.}
where 
\begin{equation}
\mathcal{M}(q_1, q_2)=\underbrace{M_{\delta_1}(q_1,q_2)^\frac{1}{2}\cdots M_{\delta_{\chi(X)}}(q_1, q_2)^\frac{1}{2}}_{\chi(X)}
\end{equation}
and $M_\delta(q_1, q_2)$ is the refined MacMahon functions defined by
\begin{equation}
M_\delta(q_1, q_2)=\prod_{i,j=1}^\infty(1-q_1^{i-\frac{1}{2}+\frac{\delta}{2}}q_2^{j-\frac{1}{2}-\frac{\delta}{2}})^{-1}, 
\label{eq11}
\end{equation}
 $\chi(X)$ is the Euler characteristic of the Calabi-Yau threefold $X$, and $\delta_i$ are constants related to how we setup the weight when we count the 3D Young diagrams \cite{Okounkov2007a, Dimofte2010a}. Thus we have the following equality:
\begin{equation}
{\cal Z}_{Fock}={\cal Z}_{top}^{ref}(q_1, q_2, Q){\cal Z}_{top}^{ref}(q_1, q_2, Q^{-1}). \label{eq9}
\end{equation}

Hence we may conjecture the following formula:
\begin{equation}
{\cal Z}_{BPS}^{ref}(chamber)={\cal Z}_{Fock}|_{chamber}={\cal Z}_{top}^{ref}(q_1, q_2, Q){\cal Z}_{top}^{ref}(q_1, q_2, Q^{-1})|_{chamber}. \label{eq10}
\end{equation}
We call it the \emph{refined} reminiscence of the OSV formula \footnote{We owe this clarification to M. Yamazaki.}. Here we need to clarify the meaning of a \emph{chamber}. The chambers here are supposed to be the same as the chambers in the unrefined case. Since the refined wall-crossing formulas are dealing with the Poincar\'{e} polynomial of the BPS states moduli space, the wall of the unrefined BPS states partition function must be the wall of the refined BPS partition. But the question here is if there exist some invisible walls beyond the walls in the unrefined case? The answer is negative. In other words, if the \emph{refined} reminiscence of the OSV formula (\ref{eq10}) is true then there do not exist any invisible walls, since if the chamber of the Euler characteristic is fixed then the right hand side of the formula (\ref{eq10}) is fixed. It implies that if there exists an invisible wall splitting the chamber into two smaller chambers, then the right hand side of the formula (\ref{eq10}) can NOT distinguish these two smaller chambers, which implies that the expression of the left hand side of  the formula (\ref{eq10}) is exactly the same in these two smaller chambers.

\section{Free fermions}
\label{sec:fermion}

The two-dimensional complex free fermion theory has been well studied for long time \cite{Jimbo1983, Macdonald1999}. Recent ten years it has shown its power in string theory and algebraic geometry \cite{Okounkov2003,  Iqbal2008, Sulkowski2007, Dijkgraaf2007, Dijkgraaf2008, Young2008}. In this section we will briefly review the vertex operators in the two-dimensional complex free fermion theory. 

We know that a 3D Young diagram can be split into some 2D Young diagrams by slicing diagonally. The partitions corresponding to these 2D Young diagrams satisfy the interlacing condition \cite{Okounkov2007b}. For each 2D Young diagram we can associate the following fermion states
\begin{equation}
  \label{eq:3}
  |\mu\rangle=\prod_{i=1}^{d(\mu)}\psi_{-a_i-\frac12}^*\psi_{-b_i-\frac12}|0\rangle, \quad \quad \quad (a_i=\mu_i-i, \quad b_i=\mu_i^t-i)
\end{equation}
where $a_i, b_i$ are called the \emph{Frobenius numbers} of the 2D Young diagram \cite{Andrews1984}, $\mu=(\mu_1, \mu_2, \cdots, \mu_\ell)$ is the partition corresponding to the 2D Young diagram, $d(\mu)$ is the number of boxes along the diagonal of the 2D Young diagram and $\psi$ and $\psi^*$ are the mode operators of the complex free fermion fields
\begin{equation}
  \label{eq:4}
  \psi(z)=\sum_{k\in {\mathbb Z}}\frac{\psi_{k+\frac12}}{z^{k+1}}, \quad \quad  \psi^*(z)=\sum_{k\in {\mathbb Z}}\frac{\psi^*_{k+\frac12}}{z^{k+1}},  
\quad\quad \{\psi_{k+\frac12}, \psi^*_{-\ell-\frac12} \}=\delta_{k\ell}.
\end{equation}
The bosonized field $:\psi(z)\psi^*(z):$ has a mode expansion 
$\sum_{n\in {\mathbb Z}}\frac{\alpha_n}{z^{n+1}}$ where the mode operators satisfy the Heisenberg algebra $[\alpha_m, \alpha_{-n}]=n\delta_{m,n}$. The vertex operators are defined as
\begin{equation}
  \label{eq:6}
\Gamma_\pm(x)=e^{\sum_{n>0}\frac{x^n}{n}\alpha_{\pm n}}, \quad\quad \Gamma'_\pm(x)=e^{\sum_{n>0}(-1)^{n-1}\frac{x^n}{n}\alpha_{\pm n}}.
\end{equation}
The actions of $\Gamma_\pm(x)$ and $\Gamma'_\pm(x)$ on $|\mu\rangle$ are as follows:
\begin{eqnarray}
\Gamma_-(x)|\mu\rangle=\sum_{\lambda\succ\mu}x^{|\lambda|-|\mu|}|\lambda\rangle, 
&\quad\quad&  \Gamma_+(x)|\mu\rangle=\sum_{\lambda\prec\mu}x^{|\mu|-|\lambda|}|\lambda\rangle, \\
\Gamma_-'(x)|\mu\rangle=\sum_{\lambda^t\succ\mu^t}x^{|\lambda|-|\mu|}|\lambda\rangle, 
&\quad\quad&  \Gamma_+'(x)|\mu\rangle=\sum_{\lambda^t\prec\mu^t}x^{|\mu|-|\lambda|}|\lambda\rangle, 
\end{eqnarray}
where the interlacing relation between partitions is defined by 
\begin{equation}
\lambda\succ\mu\quad \Longleftrightarrow \quad\lambda_1\geq \mu_1\geq\lambda_2\geq \mu_2\geq \cdots,
\end{equation}
and $\lambda^t$ means the partition corresponding to the 2D Young diagram conjugated to $\lambda$. Using the algebra of $\alpha_n$ these operators satisfy the commutation relations
\begin{eqnarray}
  \label{eq:5}
 && \Gamma_+(x)\Gamma_-(y)=\frac{1}{1-xy}\Gamma_-(y)\Gamma_+(x),\\
 &&\Gamma'_+(x)\Gamma'_-(y)=\frac{1}{1-xy}\Gamma'_-(y)\Gamma'_+(x),\\  
 &&  \Gamma'_+(x)\Gamma_-(y)=(1+xy)\Gamma_-(y)\Gamma'_+(x),\\
 &&  \Gamma_+(x)\Gamma'_-(y)=(1+xy)\Gamma'_-(y)\Gamma_+(x).
\end{eqnarray}
In the refined case, the relation between 2D partition and free fermion is preserved. But for different slices the vertex operators have different argument. 
\begin{figure}[ht]
  \centering
  \includegraphics[width=0.2\textwidth]{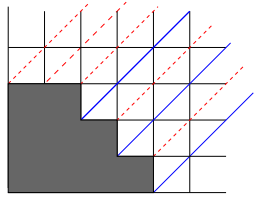}
  \caption{The red (dotted) lines denote the left or right moving of the slices, and the blue (solid) lines denote the up or down moving of the slices. }
\end{figure}

We also introduce the generalized operators for different slices and colors. $\widehat {\cal Q}_{g}$ can be $\widehat Q_{01}, \widehat Q_{02}$ for different slices and  $\widehat Q$ for K\"ahler parameters.  These operators act on $|\mu\rangle$ as follows:

\begin{equation}
\widehat {\cal Q}_{g}|\mu\rangle :=q_g^{|\mu|}|\mu\rangle.
\end{equation}

The commutation relations between $\widehat {\cal Q}_{g}$ and $\Gamma_\pm(x)$ are given by following formulas:

\begin{eqnarray}
  \label{eq:2}
&&  \widehat {\cal Q}_{g}^{-1}\Gamma_+(x)\widehat {\cal Q}_{g}=\Gamma_+(q_{g}x),\quad\quad\widehat {\cal Q}_{g}^{-1}\Gamma'_+(x)\widehat {\cal Q}_{g}=\Gamma'_+(q_{g}x), \\
&& \widehat {\cal Q}_{g}\Gamma_-(x)\widehat {\cal Q}^{-1}_{g}=\Gamma_-(q_{g}x), \quad \quad\widehat {\cal Q}_{g}\Gamma'_-(x)\widehat {\cal Q}^{-1}_{g}=\Gamma_-(q_{g}x).
\end{eqnarray}

\section{Revisiting \texorpdfstring{$\C^3$}{C3}}
\label{sec:C3}

In this section, we will reconsider the crystal corresponding to  $\C^3$ and show how to use the vertex operators in section \ref{sec:fermion} to reproduce the refined MacMahon function (\ref{eq11}). 

\begin{figure}[ht]
\centering
\includegraphics[scale=0.2]{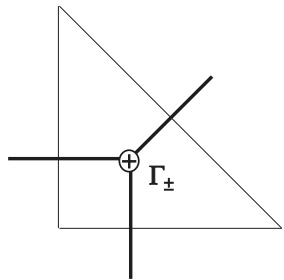}%
\caption{Toric diagram for $\C^3$ copied from \cite{Sulkowski:2009rw}}%
\label{toricC3}
\end{figure}

\begin{figure}%
\centering
\includegraphics[scale=0.2]{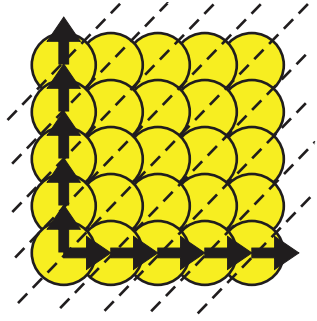}%
\caption{The crystal corresponding to $\C^3$ copied from \cite{Sulkowski:2009rw}}%
\label{crystalC3}%
\end{figure}

Figure \ref{toricC3} shows the toric diagram of $\C^3$. The crystal corresponding to $\C^3$ is shown in Figure \ref{crystalC3}.

\begin{figure}[ht]
\centering
\psfrag{Ga+}[][][0.8]{$\Gamma_+$} \psfrag{Gap+}[][][0.8]{$\Gamma'_+$}\psfrag{Ga-}[][][0.8]{$\Gamma_-$}\psfrag{Gap-}[][][0.8]{$\Gamma'_-$}
\includegraphics[width=0.7\textwidth]{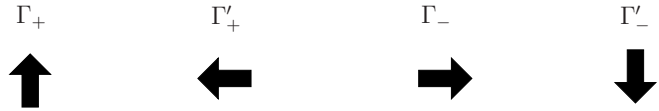}
\caption{Arrows copied from \cite{Sulkowski:2009rw}}
\label{arrow}
\end{figure}

Figure \ref{arrow} shows how we associate the arrows with the vertex operator $\Gamma_\pm, \Gamma_\pm'$ defined in section \ref{sec:fermion}.

\begin{figure}[h]%
\centering
\includegraphics[scale=0.2]{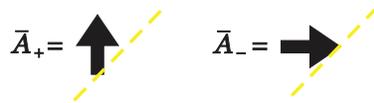}%
\caption{The relation between $A_\pm$ and the arrows}%
\label{arrow2}%
\end{figure}

Now we define
\begin{eqnarray}
  \label{eq:1}
&&  \overline A_+(x):=\widehat Q_{01}^{\frac12-\frac\delta2}\Gamma_+(x) \widehat Q_{01}^{\frac12+\frac\delta2}=\widehat Q_{01}\Gamma_+(xq_1^{\frac12+\frac\delta2}), \\
&&  \overline A_-(x):=\widehat Q_{02}^{\frac12-\frac\delta2}\Gamma_-(x) \widehat Q_{02}^{\frac12+\frac\delta2}=\Gamma_-(xq_2^{\frac12-\frac\delta2}) \widehat Q_{02}.
\end{eqnarray}
According to Figure \ref{arrow}, it is not difficult to find the relation, shown in Figure \ref{arrow2}, between $\overline{A}_\pm$ and the arrows. 

Hence we have 
\begin{eqnarray}
  \mathcal{Z}_{Fock}=\mathcal{Z}_{BPS}^{ref}&=&\langle0| \overline A_+(1) \cdots \overline A_+(1) \overline A_-(1) \cdots \overline A_-(1)|0\rangle \nonumber\\
  &=&M_\delta(q_1, q_2)\nonumber\\
  &=&(\mathcal{Z}_{top}^{ref})^2,
\end{eqnarray}
where 
\begin{equation}
  \label{eq:7}
\mathcal{Z}_{top}^{ref}=M_\delta(q_1, q_2)^{\frac{\chi(\C^3)}{2}}=M_\delta(q_1, q_2)^{\frac{1}{2}}.  
\end{equation}

\section{Resolved conifold}
\label{sec:conifold}

In this section, we will show how to use the vertex operators introduced in section \ref{sec:fermion} to reproduce the refined BPS states partition function and how to use the vertex operators to reproduce the refined BPS states partition function after flopping.

\subsection{The refined BPS states partition function and vertex operators}

\begin{figure}[h]%
\centering
\includegraphics[scale=0.2]{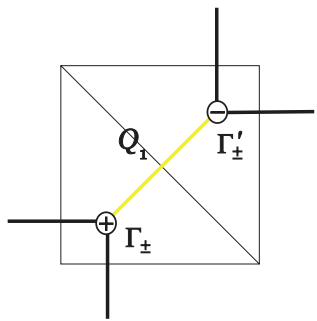}%
\caption{The toric diagram of the conifold copied from \cite{Sulkowski:2009rw}}%
\label{conifoldtoric}%
\end{figure}

The toric diagram of the conifold is shown in figure \ref{conifoldtoric}. 

\begin{figure}[h]%
\centering
\includegraphics[scale=0.2]{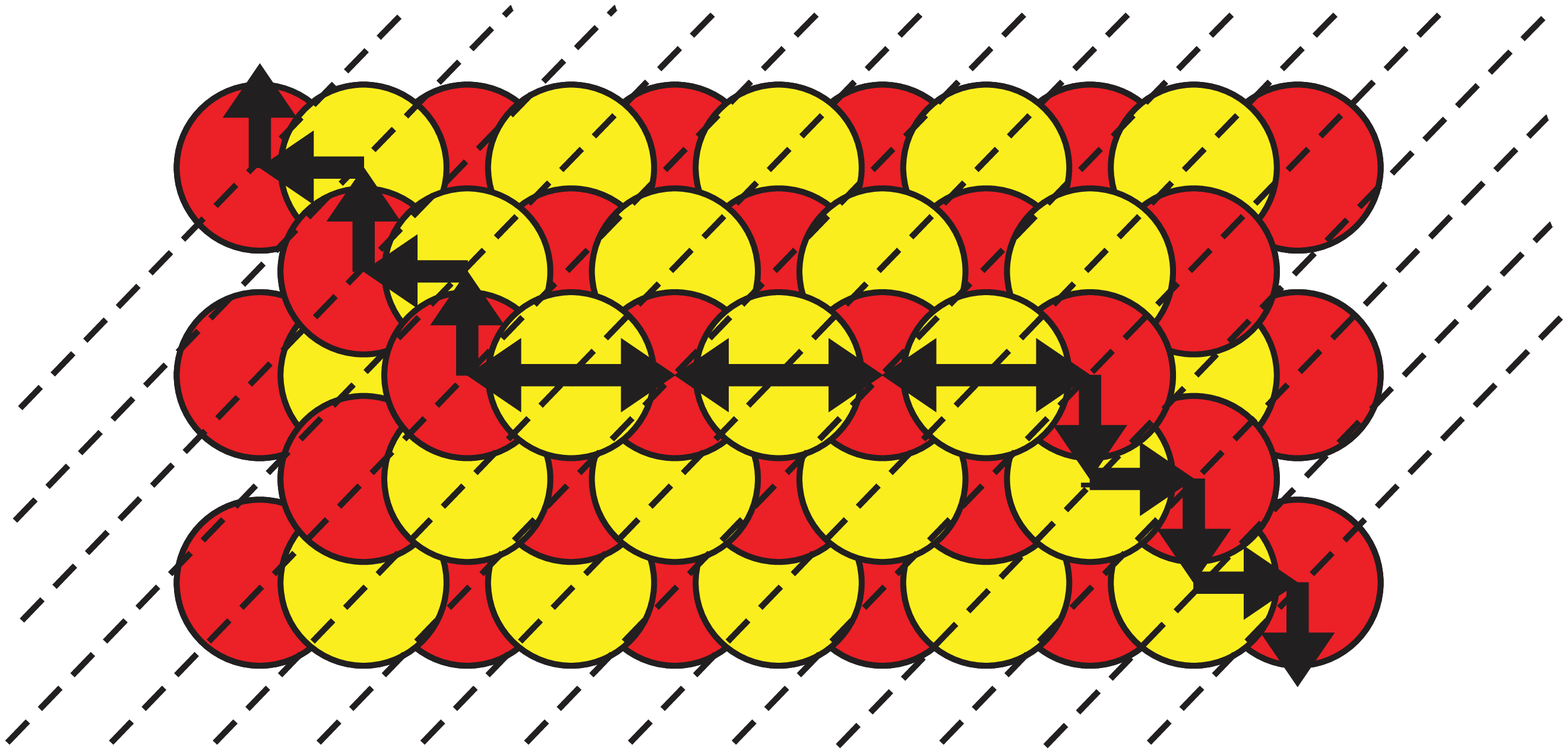}%
\caption{The crystal in the chamber $(2<B<3, R>0)$ }%
\label{coni1}%
\end{figure}

\begin{figure}[h]%
\centering
\includegraphics[scale=0.2]{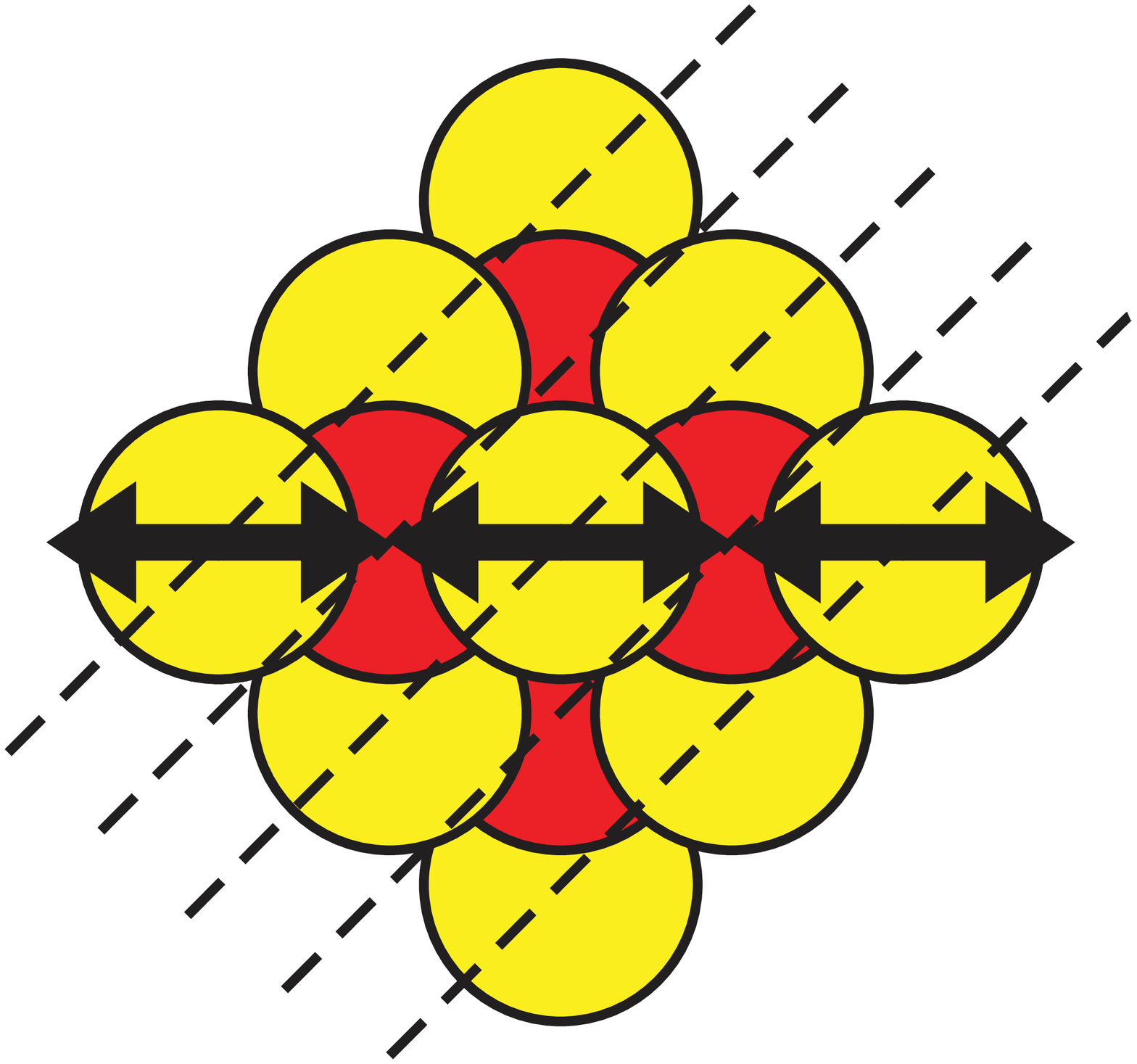}%
\caption{The crystal in the chamber $(2<B<3, R<0)$ }%
\label{coni2}%
\end{figure}

It is well known that for the unrefined BPS states partition function, we may associate a crystal in each chamber whose partition function of melting is exactly the same as the unrefined BPS states partition in that chamber \cite{Szendroi2007, Young2007, Sulkowski:2009rw}. The picture of crystals in different chambers are figure \ref{coni1} and \ref{coni2}.


\begin{figure}[h]
\centering
\psfrag{Ga+}[][][0.7]{$\Gamma_+\left[q_1^{k-\frac12}(-Q)^{\frac12}\right]$} 
\psfrag{Gap+}[][][0.7]{$\Gamma'_+\left[q_1^{k+n-\frac12}(-Q)^{-\frac12}\right]$}
\psfrag{Ga-}[][][0.7]{$\Gamma_-\left[q_2^{k+n-\frac12}(-Q)^{-\frac12}\right]$}
\psfrag{Gap-}[][][0.7]{$\Gamma'_-\left[q_2^{k-\frac12}(-Q)^{\frac12}\right]$}
\includegraphics[width=0.8\textwidth]{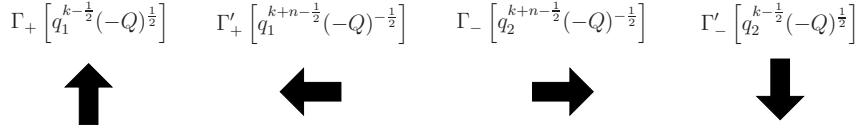}
\caption{Arrow diagrams for chamber $n$ of the conifold}
\label{fig:arrow_conifold}
\end{figure}

Now let us consider the refined BPS states partition function. Like the $\C^3$ case, we also associate the arrows in figure \ref{coni1} and \ref{coni2} with the vertex operators via figure \ref{arrow}. The detailed correspondence of the vertex operators and arrows is shown in figure \ref{fig:arrow_conifold}.

If we define $\overline{A}_{\pm}(x)$ by
\begin{eqnarray}
\overline{A}_+(x)&:=& \widehat{Q}_{01}^{\frac{1}{2}}\Gamma_+(x)\widehat{Q}_1\Gamma'_+(x)\widehat Q_{01}^{\frac{1}{2}},\\
 A_+(x)&:=& (\widehat Q_1\widehat Q_{01})^{-1}\overline A_+(x) 
=\Gamma_+\left[(-Q)^{\frac{1}{2}}q_1^{\frac{1}{2}}x\right]\Gamma'_+\left[(-Q)^{-\frac{1}{2}}q_1^{\frac{1}{2}}x\right].\\
\overline{A}_- (x)&:=& \widehat{Q}_{02}^{\frac{1}{2}}\Gamma_-(x)\widehat Q_1\Gamma'_-(x)\widehat Q_{02}^{\frac{1}{2}}\\
A_-(x)&:=&\overline A_-(x)(\widehat Q_1\widehat Q_{02})^{-1},
=\Gamma_-\left[(-Q)^{-{\frac{1}{2}}}q_2^{\frac{1}{2}}x\right]\Gamma'_-\left[(-Q)^{{\frac{1}{2}}}q_2^{\frac{1}{2}}x\right],
\end{eqnarray}
where $q_1, q_2, Q$ are defined by 
\begin{eqnarray}
q_1&:=&Q_{01}Q_1,\\
q_2&:=&Q_{02}Q_1,\\
Q&:=& -Q_1.
\end{eqnarray}
If we define states $\langle\Omega_+|$ and $|\Omega_-\rangle$ by 
\begin{eqnarray}
\langle\Omega_+|&:=&\langle0|\overline A_+(1)\cdots\overline A_+(1),\\
|\Omega_-\rangle&:=&\overline A_-(1)\cdots\overline A_-(1)|0\rangle,
\end{eqnarray}
then we can show that 
\begin{eqnarray}
\mathcal{Z}_{NCDT}^{ref}&=&\langle\Omega_+|\Omega_-\rangle \\
&=& M_{\delta=0}(q_1, q_2)^2\prod_{i,j=1}^\infty(1-q_1^{i-\frac12}q_2^{j-\frac12}Q)\prod_{i,j=1}^\infty(1-q_1^{i-\frac12}q_2^{j-\frac12}Q^{-1})\\
&=& M_{\delta=0}(q_1,q_2)^2(\mathcal{Z}_{top}^{ref}(q_1, q_2, Q))(\mathcal{Z}_{top}^{ref}(q_1, q_2, Q^{-1})),
\end{eqnarray}
which is exactly the same as the result in \cite{Dimofte2010a}.

If we are in the chamber $(R>0, 0<n<B<n+1)$, we find that the refined BPS states partition function can be written in the following formula by using the vertex operators: 
\begin{eqnarray}
\mathcal{Z}_{BPS}^{ref}&=&\langle0|\prod_{k=1}^\infty \Gamma_+\left[q_1^{k-\frac{1}{2}}(-Q)^{\frac{1}{2}}\right]\Gamma_+'\left[q_1^{k+n-\frac{1}{2}}(-Q)^{-\frac{1}{2}}\right]\cdot \nonumber\\
& & \Gamma_-\left[q_2^{\frac{1}{2}}(-Q)^{-\frac{1}{2}}\right]\Gamma_+'\left[ q_1^{n-\frac{1}{2}}(-Q)^{-\frac{1}{2}}\right] \Gamma_-\left[q_2^{\frac{3}{2}}(-Q)^{-\frac{1}{2}}\right]\Gamma_+'\left[ q_1^{n-\frac{3}{2}}(-Q)^{-\frac{1}{2}}\right]\cdots \nonumber\\
& & \Gamma_-\left[q_2^{n-\frac{1}{2}}(-Q)^{-\frac{1}{2}}\right]\Gamma_+'\left[ q_1^{\frac{1}{2}}(-Q)^{-\frac{1}{2}}\right]\prod_{k=1}^\infty\Gamma_-\left[q_2^{k+n-\frac{1}{2}}(-Q)^{-\frac{1}{2}}\right]\Gamma_-'\left[q_2^{k-\frac{1}{2}}(-Q)^{\frac{1}{2}}\right]|0\rangle,\nonumber\\
&=& M_{\delta=0}(q_1,q_2)^2\prod_{i,j=1}^\infty(1-q_1^{i-\frac12}q_2^{j-\frac12}Q)\prod_{\substack{i+j>n+1\\
i,j\geq1}}^\infty(1-q_1^{i-\frac12}q_2^{j-\frac12}Q^{-1})\nonumber\\
&=& M_{\delta=0}(q_1,q_2)^2(\mathcal{Z}_{top}^{ref}(q_1, q_2, Q))(\mathcal{Z}_{top}^{ref}(q_1, q_2, Q^{-1}))|_{(R>0, 0<n<B<n+1)},
\end{eqnarray}
which is exactly the same as the results in \cite{Dimofte2010a}.

For the chamber $(R<0, 0<B<1)$, like in \cite{Sulkowski:2009rw}, we also have
\begin{equation}
\mathcal{Z}^{ref}_{BPS}=\langle0|0\rangle=1.
\end{equation}

For the chamber $(R<0, 0<n<B<n+1)$, using the vertex operators the refined BPS states partition function can be written as 
\begin{eqnarray}
\mathcal{Z}^{ref}_{BPS}&=&\langle0|\Gamma_+'\left[ q_1^{\frac{1}{2}}(-Q)^{-\frac{1}{2}}\right]\Gamma_-\left[q_2^{n-\frac{1}{2}}(-Q)^{-\frac{1}{2}}\right]\cdots\Gamma_+'\left[ q_1^{n-\frac{1}{2}}(-Q)^{-\frac{1}{2}}\right] \Gamma_-\left[q_2^{\frac{1}{2}}(-Q)^{-\frac{1}{2}}\right]|0\rangle\nonumber\\
&=& (1-q_1^{\frac{1}{2}}q_2^{n-\frac{1}{2}}Q^{-1})\cdots (1-q_1^{n-\frac{1}{2}}q_2^{\frac{1}{2}}Q^{-1}).
\end{eqnarray}
If $q_1=q_2=q$, then all the above results will return to the unrefined case.

\subsection{Flop}

\begin{figure}[h]
  \centering
 \input{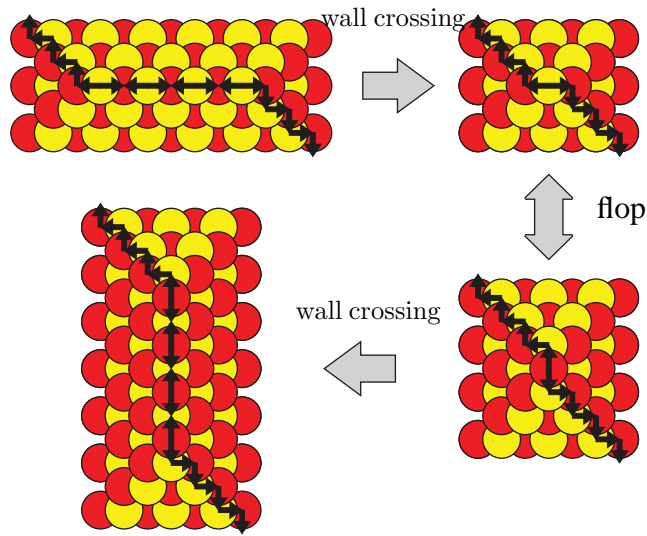}
  \caption{flop diagram}
  \label{fig:flop}
\end{figure}
According to \cite{Sulkowski:2009rw}, the effect of the flop on the conifold on the crystals is shown in figure \ref{fig:flop}.

We know that the effect of the flop is changing $Q$ to $Q^{-1}$ \cite{Szendroi2007, Taki2008}. So we propose the following correspondence between arrows and vertex operators: 

\begin{figure}[ht]
\centering
\psfrag{Ga+}[][][0.7]{$\Gamma_+\left[q_1^{k-n-\frac12}(-Q)^{\frac12}\right]$} 
\psfrag{Gap+}[][][0.7]{$\Gamma'_+\left[q_1^{k-\frac12}(-Q)^{-\frac12}\right]$}
\psfrag{Ga-}[][][0.7]{$\Gamma_-\left[q_2^{k-\frac12}(-Q)^{-\frac12}\right]$}
\psfrag{Gap-}[][][0.7]{$\Gamma'_-\left[q_2^{k-n-\frac12}(-Q)^{\frac12}\right]$}
\includegraphics[width=0.8\textwidth]{arrow}
\caption{Arrow diagrams for the chamber $n$ of the flopped conifold}
\label{fig:arrow_coniflop}
\end{figure}

For the chamber $(R>0, n-1<B<n\leq 0) $
\begin{eqnarray}
\mathcal{Z}^{ref}_{BPS}&=&\langle0|\prod_{k=1}^\infty \Gamma_+'\left[q_1^{k-\frac{1}{2}}(-Q)^{-\frac{1}{2}}\right]\Gamma_+\left[q_1^{k-n-\frac{1}{2}}(-Q)^{\frac{1}{2}}\right]\cdot \nonumber\\
& & \Gamma_-'\left[q_2^{\frac{1}{2}}(-Q)^{\frac{1}{2}}\right]\Gamma_+\left[ q_1^{-n-\frac{1}{2}}(-Q)^{\frac{1}{2}}\right] \Gamma_-'\left[q_2^{\frac{3}{2}}Q^{\frac{1}{2}}\right]\Gamma_+\left[ q_1^{-n-\frac{3}{2}}(-Q)^{\frac{1}{2}}\right]\cdots\nonumber\\
& & \Gamma_-'\left[q_2^{-n-\frac{1}{2}}(-Q)^{\frac{1}{2}}\right]\Gamma_+\left[ q_1^{\frac{1}{2}}(-Q)^{\frac{1}{2}}\right]\prod_{k=1}^\infty\Gamma_-'\left[q_2^{k-n-\frac{1}{2}}(-Q)^{\frac{1}{2}}\right]\Gamma_-\left[q_2^{k-\frac{1}{2}}(-Q)^{-\frac{1}{2}}\right]|0\rangle\nonumber\\
&=& M_{\delta=0}(q_1, q_2)^2\prod_{i,j=1}^\infty(1-q_1^{i-\frac{1}{2}}q_2^{j-\frac{1}{2}}Q^{-1})\prod_{\substack{i+j> -n+1\\
i,j\geq 1}}^\infty(1-q_1^{i-\frac{1}{2}}q_2^{j-\frac{1}{2}}Q).
\end{eqnarray}
The refined NCDT after flopping is corresponding to the $n=0$ case. 

For the chamber $(R<0, -1<B<0)$, we have
\begin{equation}
\mathcal{Z}^{ref}_{BPS}=\langle0|0\rangle=1.
\end{equation}

For the chamber $R<0, n-1<B<n<0$, we have
\begin{eqnarray}
\mathcal{Z}^{ref}_{BPS}&=&\langle0|\Gamma_+\left[ q_1^{\frac{1}{2}}(-Q)^{\frac{1}{2}}\right]\Gamma_-'\left[q_2^{-n-\frac{1}{2}}(-Q)^{\frac{1}{2}}\right]\cdots\Gamma_+\left[ q_1^{-n-\frac{1}{2}}(-Q)^{\frac{1}{2}}\right] \Gamma_-'\left[q_2^{\frac{1}{2}}Q^{\frac{1}{2}}\right]|0\rangle\nonumber\\
&=& (1-q_1^{\frac{1}{2}}q_2^{-n-\frac{1}{2}}Q)\cdots (1-q_1^{-n-\frac{1}{2}}q_2^{\frac{1}{2}}Q).
\end{eqnarray}


\section{\texorpdfstring{$\mathcal{O}(-2)\oplus \mathcal{O}\rightarrow \mathbb{P}^1$}{O(-2)+O}}
\label{sec:Ominus2}

In this section, we will see how to use the vertex operators to get the refined BPS states partition function and to check the correctness of the \emph{refined} reminiscence of the OSV formula (\ref{eq10}).

\begin{figure}[h]%
\centering
\includegraphics[scale=0.2]{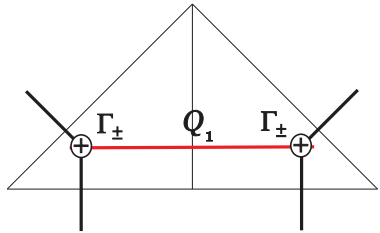}%
\caption{The toric diagram of $\mathcal{O}(-2)\oplus \mathcal{O}\rightarrow \mathbb{P}^1$ copied from \cite{Sulkowski:2009rw}}%
\label{O-2toric}%
\end{figure}

\begin{figure}[h]
  \centering
  \includegraphics[width=0.35\textwidth]{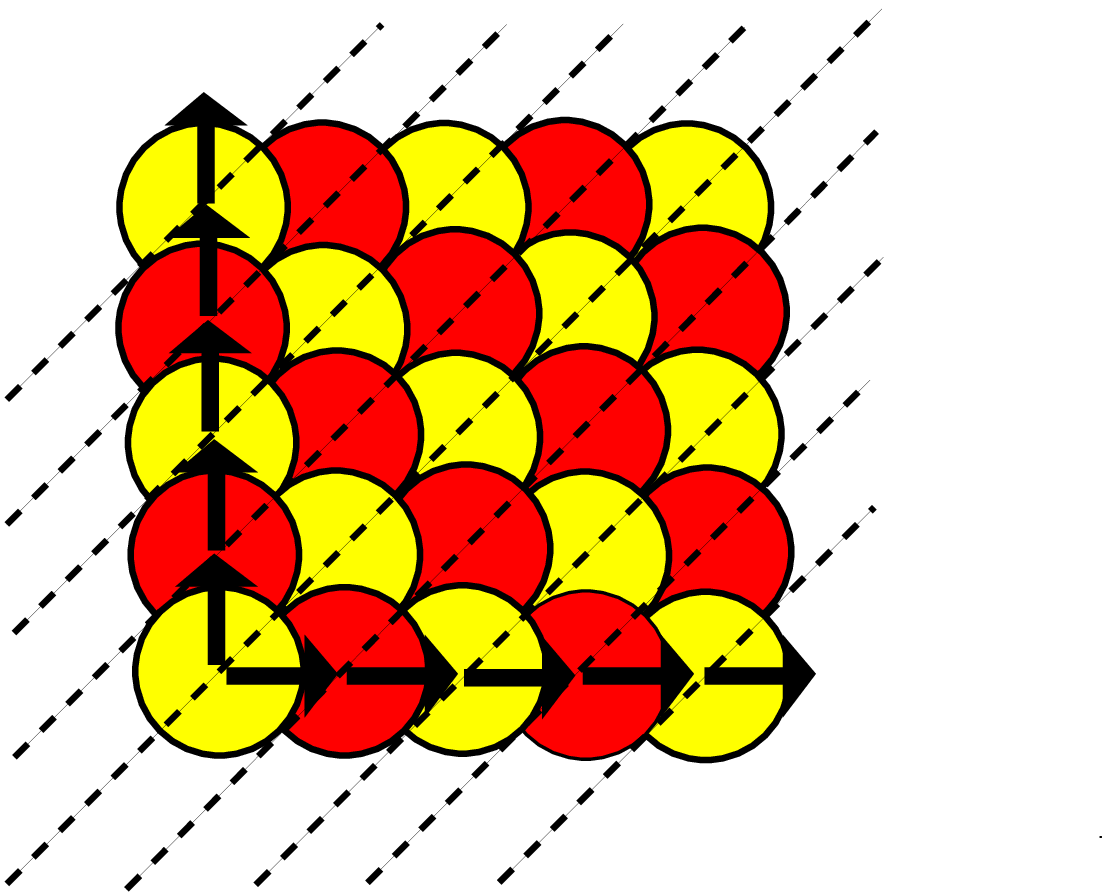}
  \caption{The crystal for the NCDT chamber of ${\cal O}(-2)\oplus{\cal O}\rightarrow \mathbb{P}^1$ }
  \label{fig:crystal_O2}
\end{figure}

\begin{figure}[h]%
\centering
\includegraphics[scale=0.2]{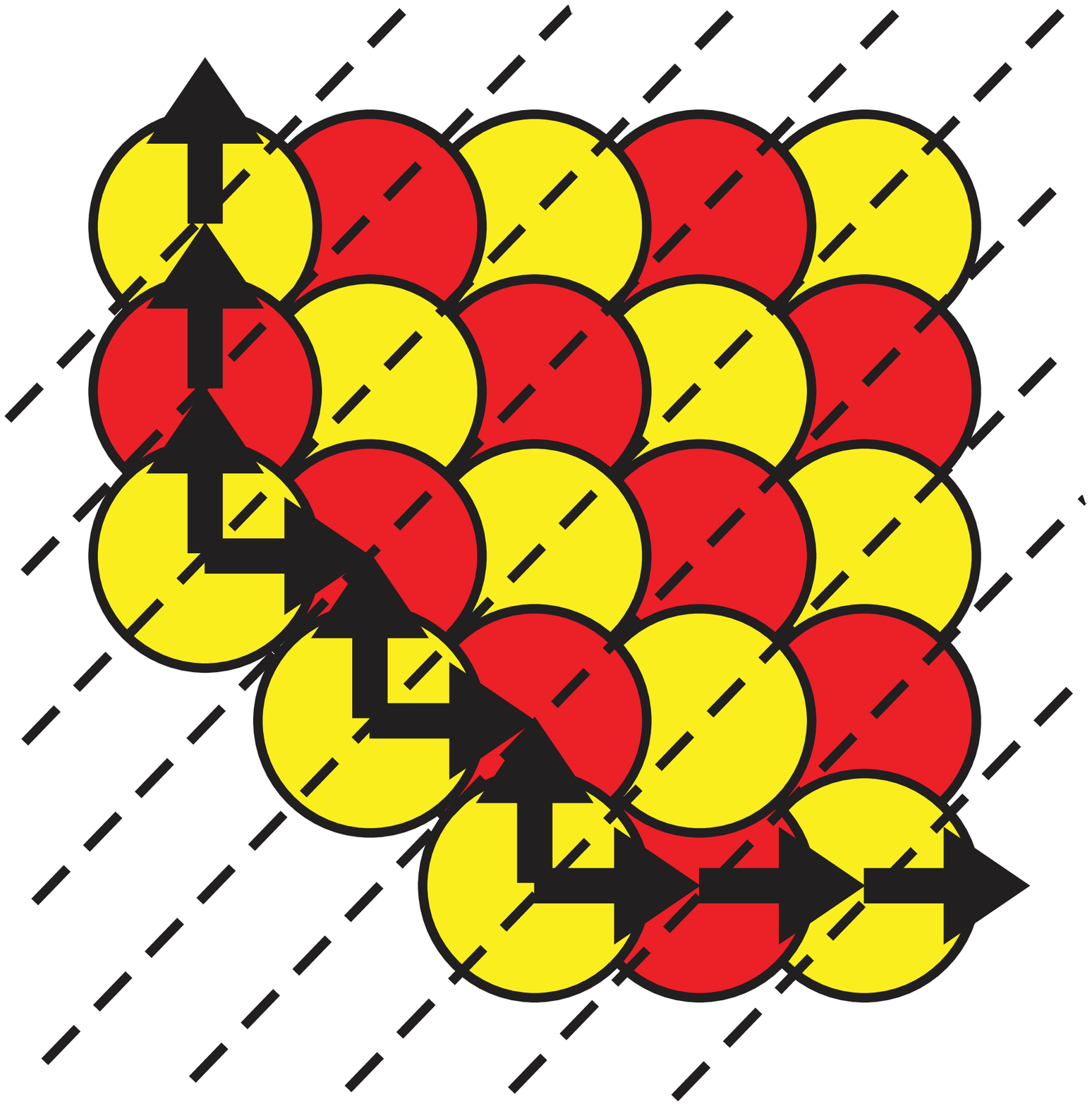}%
\caption{The crystal in the chamber $(R>0, 2<B<3)$ copied from \cite{Sulkowski:2009rw}}%
\label{O-2crystal1}%
\end{figure}

\begin{figure}[ht]%
\centering
\includegraphics[scale=0.2]{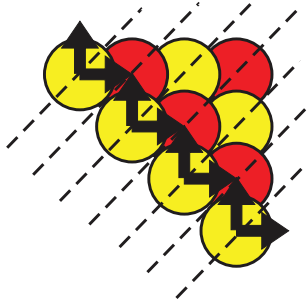}%
\caption{The crystal in the chamber $(R<0, 4<B<5)$ copied from \cite{Sulkowski:2009rw}}%
\label{O-2crystal2}%
\end{figure}

\begin{figure}[h]
\centering
\psfrag{Ga+}[][][0.7]{$\Gamma_-\left[q_2^{k-1}(-Q)^{\frac12}\right]$}
\psfrag{Gap+}[][][0.7]{$\Gamma_-\left[q_2^{k+n-1}(-Q)^{-\frac12}\right]$}
\psfrag{Ga-}[][][0.7]{$\Gamma_+\left[q_1^{k}(-Q)^{\frac12}\right]$} 
\psfrag{Gap-}[][][0.7]{$\Gamma_+\left[q_1^{k+n}(-Q)^{-\frac12}\right]$}
\includegraphics[width=0.8\textwidth]{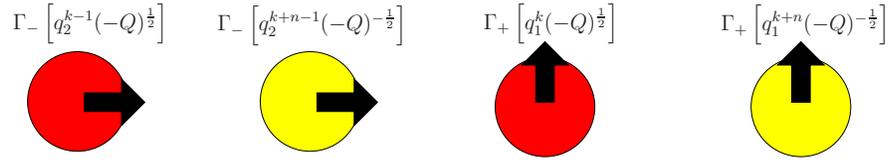}
\caption{Arrow diagrams for chamber $n$ of ${\cal O}(-2)\oplus{\cal O}\rightarrow \mathbb{P}^1$. We use the red and yellow stones to distinguish those vertex operators which has $(-Q)^{\frac12}$ 
or $(-Q)^{-\frac12}$ argument}
\label{fig:arrow_Ominus2}
\end{figure}

The toric diagram of $\mathcal{O}(-2)\oplus \mathcal{O}\rightarrow \mathbb{P}^1$ is shown in figure \ref{O-2toric}. 
Figure \ref{fig:crystal_O2} shows the crystal of the NCDT chamber in the unrefined case. Figure \ref{O-2crystal1} and \ref{O-2crystal2} show the crystals in the unrefined case in some nontrivial chambers.
Like the $\C^3$ and resolved conifold cases, we also associate some vertex operators with the arrows (figure \ref{fig:arrow_Ominus2}) in figure \ref{fig:crystal_O2}, \ref{O-2crystal1}, \ref{O-2crystal2}. 

First, let us see the NCDT chamber $(R>0, 0<B<1)$. In this chamber the crystal diagram is shown in figure \ref{fig:crystal_O2}. According to the relation between the arrows and vertex operators shown in figure \ref{fig:arrow_Ominus2}, we may find that
\begin{eqnarray}
\mathcal{Z}_{NCDT}^{ref}&=&\langle0|\prod_{k=1}^\infty\Gamma_+\left[q_1^{k}(-Q)^{\frac12}\right]\Gamma_+\left[q_1^{k}(-Q)^{-\frac12}\right]\prod_{k=1}^\infty\Gamma_-\left[q_2^{k-1}(-Q)^{\frac12}\right]\Gamma_-\left[q_2^{k-1}(-Q)^{-\frac12}\right]|0\rangle\nonumber\\
&=&M_{\delta=1}^2(q_1,q_2)\prod_{i,j=1}^\infty(1-q_1^{i}q_2^{j-1}Q)^{-1}\prod_{i,j=1}^\infty(1-q_1^{i}q_2^{j-1}Q^{-1})^{-1}\nonumber\\
&=&M_{\delta=1}^2(q_1, q_2)\mathcal{Z}_{top}^{ref}(q_1, q_2, Q)\mathcal{Z}_{top}^{ref}(q_1, q_2, Q^{-1})|_{NCDT chamber}
\label{eq:}
\end{eqnarray}

We may also define $\overline A_\pm(x)$ by
\begin{eqnarray}
&&\overline A_+(x):=\widehat Q_1^{-\frac{1}{2}}\Gamma_+(x)\widehat Q_1\Gamma_+(x)\widehat Q_{01}\widehat Q_1^{\frac{1}{2}},\\
&&A_+(x):=\overline A_+(x)(\widehat Q_1\widehat Q_{01})^{-1}=\Gamma_+\left[(-Q)^{\frac{1}{2}}x\right]\Gamma_+\left[(-Q)^{-\frac{1}{2}}x\right],\\
&&\overline A_-(x):=\widehat Q_1^{-\frac{1}{2}}\Gamma_-(x)\widehat Q_1\Gamma_-(x)\widehat Q_{02}\widehat Q_1^{\frac{1}{2}},\\
&&A_-(x):=\widehat A_-(x)(\widehat Q_1\widehat Q_{02})^{-1}= \Gamma_-\left[(-Q)^{-\frac{1}{2}}x\right]\Gamma_-\left[(-Q)^{\frac{1}{2}}x\right].
\end{eqnarray}
where $Q=-Q_1, q_1=Q_{01}Q_1, q_2=Q_{02}Q_1$.
If we define$\langle \Omega_+|$ and $|\Omega_-\rangle$ by

\begin{eqnarray}
\langle\Omega_+|&:=&\langle0|\overline A_+(1)\cdots\overline A_+(1),\\
|\Omega_-\rangle&:=&\overline A_-(1)\cdots\overline A_-(1)|0\rangle,
\end{eqnarray}
then we can show that 
\begin{equation}
\mathcal{Z}_{NCDT}^{ref}=\langle\Omega_+|\Omega_-\rangle\nonumber\\
\end{equation}


For the chamber $(R>0, 0<n<B<n+1) $, according to the correspondence between arrows and vertex operators (figure \ref{fig:arrow_Ominus2}), we have the following formula:
\begin{eqnarray}
\mathcal{Z}_{BPS}^{ref}&=&\langle0|\prod_{k=1}^\infty \Gamma_+\left[q_1^{k}(-Q)^{\frac{1}{2}}\right]\Gamma_+\left[q_1^{k+n}(-Q)^{-\frac{1}{2}}\right]\cdot \nonumber\\
& & \Gamma_-\left[(-Q)^{-\frac{1}{2}}\right]\Gamma_+\left[ q_1^{n}(-Q)^{-\frac{1}{2}}\right] \Gamma_-\left[q_2(-Q)^{-\frac{1}{2}}\right]\Gamma_+\left[ q_1^{n-1}(-Q)^{-\frac{1}{2}}\right]\cdots\nonumber\\
& &\Gamma_-\left[q_2^{n-1}(-Q)^{-\frac{1}{2}}\right]\Gamma_+\left[ q_1(-Q)^{-\frac{1}{2}}\right]\cdot \prod_{k=1}^\infty\Gamma_-\left[q_2^{k+n-1}(-Q)^{-\frac{1}{2}}\right]\Gamma_-\left[q_2^{k-1}(-Q)^{\frac{1}{2}}\right]|0\rangle\nonumber\\
&=& M_{\delta=1}^2(q_1,q_2)\prod_{i,j=1}^\infty(1-q_1^{i}q_2^{j-1}Q)^{-1}\prod_{i+j>n+1}(1-q_1^{i}q_2^{j-1}Q^{-1})^{-1}\nonumber\\
&=& M_{\delta=1}^2(q_1,q_2)\mathcal{Z}_{top}^{ref}(q_1, q_2, Q)\mathcal{Z}_{top}^{ref}(q_1, q_2, Q^{-1})|_{(R>0, 0<n<B<n+1)}
\end{eqnarray}
Similarly for the chamber $(R<0, 0<B<1)$,we also have 
\begin{equation}
\mathcal{Z}_{BPS}^{ref}=\langle0|0\rangle=1,
\end{equation}
and for the chamber $R<0, 1\leq n<B<n+1$,we also have 
\begin{eqnarray}
\mathcal{Z}_{BPS}^{ref}&=&\langle0|\Gamma_+\left[ q_1(-Q)^{-\frac{1}{2}}\right]\Gamma_-\left[q_2^{n-1}(-Q)^{-\frac{1}{2}}\right]\cdots\Gamma_+\left[ q_1^{n}(-Q)^{-\frac{1}{2}}\right] \Gamma_-\left[(-Q)^{-\frac{1}{2}}\right]|0\rangle\nonumber\\
&=&\prod_{i,j=1}^n(1-q_1^iq_2^{j-1}Q^{-1})^{-1}.
\end{eqnarray}

\section{Discussion}
\label{sec:discussion}

In this paper, we first use the M-theory to propose the so-called \emph{refined} reminiscence of the 
OSV formula connecting the \emph{refined} BPS states partition function with the \emph{refined} topological string partition function. Secondly, we produce the wall-crossing formulas of the \emph{refined} BPS states partition function by employing the vertex operators in 2-dimensional free fermions. Like in \cite{Aganagic2009}, here we also require that the Calabi-Yau threefold $X$ does NOT contain any compact four cycles. 

For the the refined reminiscence of the 
OSV formula we proposed in section \ref{sec:ref_OSV}, in the equation (\ref{eq8}) there is a factor $\mathcal{M}(q_1, q_2)$, which is related to the refined MacMahon function, appearing in the refined topological string partition function. It should be related to the D0-branes bounded to the single D6-brane. We notice that in \cite{Behrend2009}, the authors investigated the \emph{motivic} degree zero of Donaldson-Thomas generating function on $\C^3$ and some other smooth projective threefolds. In the appendix of their paper, they explained how to relate the \emph{motivic} degree zero of Donaldson-Thomas generating function on $\C^3$ to Okounkov et al.'s refined generating function in counting plane partitions in \cite{Okounkov2007a}. For example, the \emph{motivic} degree zero of Donaldson-Thomas generating function on $\C^3$ is related to the refined MacMahon function \footnote{Here we use the convention $\frac\delta2$ in equation (\ref{eq11}) as the definition of the refined MacMahon function $M_\delta(q_1,q_2)$ which is a little bit different with the definition in \cite{Behrend2009}.}  $M_{\delta=-3}(q_1,q_2)$ appearing in \cite{Okounkov2007a}. For other toric CYs, the theorem 3.7 in \cite{Behrend2009} shows how to refine the MacMahon function. The puzzle is that the refined MacMahon functions appearing in \cite{Behrend2009} are different with the refined MacMahon functions proposed in this paper. It would be very interesting to see how to reproduce the refined MacMahon functions appearing in \cite{Behrend2009} by using the method proposed in this paper. We believe that well-understanding on \cite{Behrend2009} will shed some light on this direction. 

In this paper, we only investigate the simplest three Calabi-Yau threefolds $\C^3, \mathcal{O}(-1)\oplus\mathcal{O}(-1)\rightarrow \mathbb{P}^1$, and $\mathcal{O}(-2)\oplus\mathcal{O}\rightarrow \mathbb{P}^1$. Then we may ask for other more complicated threefolds, especially those coming from the strip \cite{Iqbal2006b}, how do we use the vertex operator techniques to write down the wall-crossing formulas of the refined BPS partition functions for them? This work is in progress.  

Further, in \cite{Dimofte2010a}, the authors investigated the refined wall-crossing for the resolved conifold case by using the dimer model. We believe that using dimer model we can also get the refined wall-crossing formulas for the $\mathcal{O}(-2)\oplus\mathcal{O}\rightarrow \mathbb{P}^1$ and other cases. We are working in this direction. 

Finally, in \cite{Ooguri2010a, Szabo2010c}, the authors investigated the relation between the matrix models and the wall-crossing. For the refined wall-crossing case, it is not difficult to find that one can indeed use matrix model to reproduce the refined NCDT partition function. It would be interesting to use the matrix model to reproduce the refined wall-crossing formula; it would also be interesting to see the details of this kind of matrix model, e.g. spectral curve, etc. This work is under consideration.

\section*{Acknowledgements}

We would like to thank Prof. Bal\'{a}zs Szendr\H{o}i and Dr. Masahito Yamazaki for the valuable comments. 

H. Liu thanks Yunfeng Jiang for the valuable discussion at the final stage of this project and also thanks the hospitality and inspiring atmosphere of the 8th Simons Workshop on Mathematics and Physics and the hospitality of the theoretical physics group of Imperial College while preparing this paper. H. Liu also appreciates Prof. Jack Gegenberg for his support. The research of H. Liu is partially supported by the ``Pam and John Little Overseas Scholarship'' in the University of New Brunswick, Canada. 

J. Yang thanks the wonderful working atmosphere provided by SISSA. The experience and the friendship are most valuable for her.

\bibliographystyle{JHEP}
\bibliography{RWC}
\end{document}